\documentclass[aps,prb,twocolumn,showpacs,floatfix,superscriptaddress,amssymb,amsmath]{revtex4}
\usepackage{graphicx}
\usepackage{dcolumn}
\usepackage{bm}
\usepackage{amsmath}
\usepackage{color}
\newcommand{\dlangle}{\langle\langle}
\newcommand{\drangle}{\rangle\rangle}

\begin{document}

\title{Polarized currents in Coulomb blockade and Kondo regimes without magnetic fields}

\author{Anh T. Ngo}
\affiliation{Department of Physics and Astronomy, and Nanoscale and Quantum Phenomena Institute, Ohio University, Athens, Ohio 45701-2979}

\author{Edson Vernek}
\affiliation{Department of Physics and Astronomy, and Nanoscale and Quantum Phenomena Institute, Ohio University, Athens, Ohio 45701-2979}
\affiliation{Instituto de F\'isica - Universidade Federal de Uberl\^andia, Uberl\^andia, MG  38400-902, Brazil}

\author{Sergio E. Ulloa}
\affiliation{Department of Physics and Astronomy, and Nanoscale and Quantum Phenomena Institute, Ohio University, Athens, Ohio 45701-2979}

\date{\today}

\begin{abstract}
We present studies of the Coulomb blockade and Kondo regimes of
transport through a quantum dot connected to current leads through
\textit{spin-polarizing} quantum point contacts (QPCs). This structure,
arising from the effect of \textit{lateral} spin-orbit fields defining the QPCs,
results in spin-polarized currents even in the absence of
external magnetic fields and greatly affects the correlations in
the dot. Using equation-of-motion and numerical renormalization
group calculations we obtain the conductance and spin polarization
for this system under different parameter regimes.  We find that the system exhibits
spin-polarized conductance in both the Coulomb blockade and Kondo regimes,
{\em all in the absence of applied magnetic fields.} We analyze the role that the
spin-dependent tunneling amplitudes of the QPC play in determining
the charge and net magnetic moment in the dot.
These effects, controllable by lateral gate
voltages, may provide an alternative approach for exploring Kondo
correlations, as well as possible spin devices.
\end{abstract}

\maketitle

\section{Introduction}
Electronic transport in semiconducting nanostructures is studied
both as it has great potential applications in spintronics, \cite{5} and because of its exquisite
control of parameters, which allow insightful probes into fundamental physical phenomena.
Electrons in such systems experience
externally controlled confining environments that result in strong Coulomb interactions
with other electrons. As such, quantum dot (QD) structures
provide well-characterized and defined systems for studying quantum many-body physics.
QDs may also allow the realization of solid state quantum
computation devices as well as spintronic
semiconductor devices with unprecedented functionalities. \cite{QDref}

Manipulation of spin-polarized current sources is
crucial in spintronics. This typically requires
efficient spin injection into conventional semiconductors. The
difficulties with spin injection from ferromagnetic metal leads has
stimulated extensive efforts to produce spin polarized currents out
of unpolarized sources. \cite{Schmidt}
In this context, the Rashba
spin-orbit (RSO) coupling mechanism provides a basis for possible device
applications. \cite{Rashba}
This coupling, arising from interfacial structure asymmetries, depends on the materials used
as well as on the confinement geometry of the structures. \cite{6,9}
Most interestingly, the Rashba effect allows external
tunability, which has been studied experimentally in QDs
\cite{7} and quantum point contacts. \cite{Bird,8}

In this paper we study the electronic transport through a
quantum dot connected to polarizing quantum point contacts (QPCs) in
both the Coulomb blockade (CB) and Kondo regimes. Due to strong
spin-orbit interactions, \cite{8,9} QPCs can
exhibit spin-dependent hybridization of the QD states with the leads, {\em without}
applied magnetic fields, opening the possibility for generating
spin-polarized transport in an all-electrical setup.
These effects are
controllable by lateral gate voltages applied on QPCs, resulting in spatially
asymmetric structures, as in recent
experiments.\cite{8}
Using the
equation-of-motion technique and numerical renormalization group
(NRG) calculations we obtain the electronic Green's function, conductance and
spin polarization in different system regimes. Our results show
that both the CB and Kondo regimes exhibit non-zero
spin-polarized conductance in this system. We analyze how the spin-dependent
hybridization of the QPC modifies the charge accumulation in the dot, as well
as the density of states (spectral functions) of the system.  Interestingly,
we find that the polarizing QPCs produce spin
polarization and split DOS in the Kondo regime akin to that reported for current
injection from ferromagnetic leads, \cite{Martinek} although here it occurs for unpolarized
reservoirs.  Our theoretical studies suggest that these effects could be accessible in
experiments and result in future spintronic devices. 

The paper is organized as
follows. First, we present the main features of the current polarization in
quantum point contact systems with lateral spin-orbit, obtained by scattering matrix
methods in section \ref{QPCs}.  The next section describes the model of
a quantum dot connected to current leads via polarizing QPCs. In section \ref{EOMsect}
we use the equation-of-motion (EOM) technique to obtain the dot Green's function and discuss
numerical results in both the CB and Kondo regimes. In section
\ref{NRGsect} we revisit the problem using the numerical renormalization
group approach, which provides the most reliable information in the Kondo regime
for realistic system parameters.
The paper concludes with final remarks and summary in section \ref{summ}.

\section{Polarizing QPC} \label{QPCs}

The effective electric field in the $z$ direction that creates
a two-dimensional electron gas (2DEG) confined to the
$x$-$y$ plane results in
the well-known Rashba spin orbit interaction, \cite{Rashba,6}
\begin{eqnarray}
H_{SO}^{R}=\frac{\alpha}{\hbar} \left(\sigma_{x}P_{y}-\sigma_{y}P_{x} \right).
\end{eqnarray}
Here, $\sigma_x$ and $\sigma_y$ are Pauli matrices, and $\alpha$ is
Rashba spin orbit coupling constant, which is proportional to the
field and is therefore material and structure dependent.
Electrons
on this 2DEG entering a quantum dot, pass through QPCs defined via a
confining potential $V(x,y)$ which can be thought as made of two
parts: $U(y)+V_{b}(x,y)$, where $U(y)$ defines a hard-wall potential
of width $W$ (related in the experiment to the side-wall etching
defining the structure) outlining the overall channel structure,
while $V_{b}(x,y)$ is the potential generating the QPC barrier, and
effectively modulated by the side gate potentials of the structure.
\cite{8} We model such barrier by \cite{12}
\begin{eqnarray}
V_{b}(x,y)=\frac{1}{2}V_g\left(1+ \cos \frac{\pi x}{L_x}\right) +
\frac{1}{2}m \omega^{2} \bar{y}^2 \Theta(\bar{y})
\end{eqnarray}
with $\bar{y} = y - y_s$, and
\begin{eqnarray}
y_s=W_1 \left(1- \cos \frac{\pi x}{L_x} \right) \, ,
\end{eqnarray}
where $\Theta(x)$ is the step function, $m$ is the effective mass of
the electrons, $L_x$ is the unit length of the structure in
the $x $ direction (along the current direction) and $\omega$ is the confinement potential frequency.
 Notice
that this potential form is asymmetric in the $y$ direction, to reflect an essential
ingredient in the experiments: the QPC potential must lack $y$-reflection symmetry
in order to generate the polarizing effect along the $z$-direction. \cite{12,12b,12c}
In fact, the fields forming $V_b$ generate a spin-orbit
coupling given by  \cite{6}
\begin{eqnarray}
V_{SO}^{\beta}=-\frac{\beta}{\hbar} \nabla V_b \cdot (\mathbf{\hat{\sigma}}\times
\mathbf{\hat{P}}),
\end{eqnarray}
where $\beta$ is material-specific.  Notice that $\nabla V_b$
lies in the $x$-$y$ plane, so that the barrier
fields induce a  {\em lateral} spin-orbit coupling.
The total Hamiltonian of the QPC will then be given by,
\begin{eqnarray}
H=\frac{P_x^2+P_y^2}{2m}+H_{SO}^R+ V(x,y)+V_{SO}^{\beta}.
\end{eqnarray}

\begin{figure}[h]
\includegraphics[width=1\linewidth]{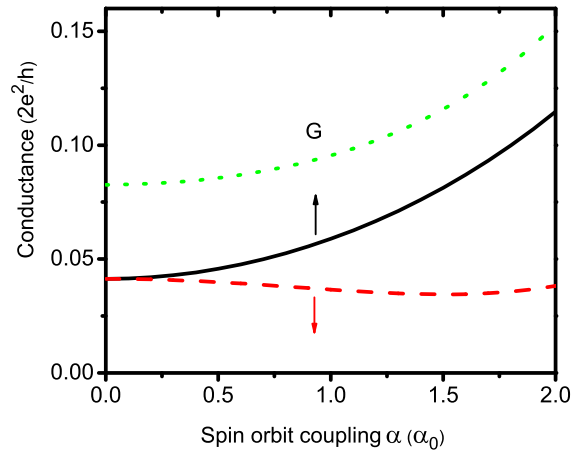}
\caption{(color online) Total and spin-dependent conductances for an asymmetric QPC as
function of Rashba spin-orbit coupling $\alpha$, obtained from a scattering matrix approach.\cite{9}
The interplay of vertical and lateral SO effects may result in large asymmetry for the
up and down spin components for realistic structure parameters, even in the tunneling regime
shown here ($G_\sigma \ll 1$).} \label{1}
\end{figure}

Using a scattering-matrix formalism to study the spin-dependent
electron transport in this QPC, \cite{11,9}
we find that a net spin-polarized conductance is produced only for $y$-asymmetric potentials. \cite{9}
We can calculate the conductance of the structure,
assuming that the SO coupling $\alpha$ and $V(x,y)$ are zero at the source and drain 2DEG reservoirs,
while both of these
terms in the Hamiltonian are turned on in the QPC region.  Typical structure parameters
in experiments can be cast in terms of characteristic length and energy scales,
$L_0=32.5 $ nm, and $E_0=3.12 $ meV, with $\alpha_0=E_0L_0=1.0 \times
10^{-12}$ eVm, a typical value of spin-orbit coupling.  Using  $W=L_x=2L_0$ and $W_1=0.6 L_0$ as
width/length of the confining potential, with $\omega=6 \times 10^{13}$s$^{-1}$,
and $\beta=0.97\times10^{-16}$m$^2$,
 gives results as shown in Fig.\ \ref{1}.  This figure shows
spin-dependent conductances as functions of Rashba
coupling $\alpha$, for a potential barrier which is near its conduction onset
(or ``pinch-off", as controlled by the value of $V_g$); arrows indicate the
results of spin up and down conductances. We see
that conductances $G_{\uparrow}$ and $G_{\downarrow}$ can be very
different from each other, even in the tunneling regime (where
each $G_\sigma \ll 1$) in which the QPC would operate to create
a quantum dot in the 2DEG.\@  We stress that for these realistic values
of structure parameters, one obtains non-zero spin polarization even
when there is {\em no external magnetic field} and the injection is
{\em unpolarized}. This interesting result can be understood
from anticrossing features in the subband energy structure in the channel region defining the QPC.
\cite{8,9} The spin mixings and avoided crossings generate spin rotation as
electrons pass through the narrow constriction of the QPC, and can generate large
values of the ratio $G_{\uparrow}/G_{\downarrow}$, even in the tunneling regime.
Two of these QPCs can then be used to define the QD and result in interesting
charging and conductance regimes, as we will see below.

\section{Quantum dot with polarizing QPCs} \label{QDmodel}

In order to address the transport through a quantum dot formed with polarizing QPCs,
we consider the single impurity Anderson model given by the following Hamiltonian:
\begin{eqnarray}
H=\sum_{\ell k\sigma}\varepsilon_{\ell k}c^\dagger_{\ell
k\sigma}c_{\ell k\sigma}+\sum_{\sigma}
\varepsilon_dc^\dagger_{d\sigma}
c_{d\sigma}+Un_{d\uparrow}n_{d\downarrow}\nonumber \\+\sum_{\ell
k\sigma}t_{\sigma}
(c^\dagger_ { d\sigma}c_{\ell k\sigma}
+c^\dagger_{\ell k\sigma}c_{d\sigma}),
\end{eqnarray}
where $c^\dagger_{d\sigma} (c_{d\sigma})$ is the creation
(annihilation) operator of an electron of spin $\sigma$ in the
dot. The quantities $\varepsilon_{\ell k}$, $\varepsilon_d$ are the
energies of the electrons in the $\ell^{\it th}$ conduction band
channels ($\ell=L,R$) and the single local energy level in the dot, respectively.
$U$ is the Coulomb repulsion between
electrons occupying the QD  with $n_{d\sigma}=c^\dagger_{d\sigma}
c_{d\sigma}$, while $t_{\sigma}$ represents the lead-QD hybridization
occurring via tunneling through the QPC, and which is assumed to be
$k$-independent. The density of states for conduction electrons in each lead is
taken to be constant,
$\rho_L(\varepsilon)=\rho_R(\varepsilon)\equiv \rho
=(1/2D)\Theta(D- |\varepsilon|)$, where $D$ is the conduction band halfwidth (hereafter taken as our energy unity).

The theoretical description of such quantum dot system, especially in the strong
correlations regime, has been greatly developed over the years. \cite{Mahan}  Techniques of note include
quantum Monte-Carlo, \cite{2} equations of motion for the Green's functions,\cite{3} and
the numerical renormalization group approach.\cite{costi}   In what follows, we explore the role that
SO interactions play on the Coulomb blockade and Kondo regimes of transport of the
QD, utilizing equations of motion and numerical renormalization group formalisms.

\subsection{Equation of motion approach and numerical results} \label{EOMsect}

To calculate the charge and conductance of the system we calculate
Green's functions (GFs), which allow us to take into account
the correlations induced by the Coulomb interaction in the QD.\@ The
retarded double-time Green's functions are defined
as ($\hbar=1$)\cite{13}
\begin{equation}
 i\dlangle A;B\drangle =\int_{-\infty}^\infty\langle
[A(\tau),B(0)]_+\rangle\Theta(\tau)e^{-i\omega \tau}d\tau,
\label{gf3}
\end{equation}
where $A$ and $B$ are generic fermionic operators, $[A,B]_+$ indicates their anticommutator and
$\langle \cdots\rangle$ indicates the thermodynamic average for
$T>0$, or the ground state expectation value for $T=0$. The GF can be obtained using
equation of motion (EOM) techniques, so that
\begin{eqnarray}\label{EOM}
 \omega\dlangle A;B\drangle =\langle [A,B]_+
\rangle+\dlangle[A,H];B\drangle, \label{gf2}
\end{eqnarray}
where $[A,B]$ represents a commutator.  Iteration of this formula
generates a hierarchy of expressions, starting with the local one-particle GF as
\begin{equation}
\left(\omega-\varepsilon_d-\sum_k \frac{\tilde
{t}_{\sigma}^2}{\omega-\varepsilon_k}\right)   \dlangle
c_{d\sigma};c^\dagger_{d\sigma} \drangle =1 +U\dlangle
c_{d\sigma}n_{d\bar{\sigma}};c^\dagger_{d\sigma} \drangle , \label{gf1}
\end{equation}
where $\tilde t_{\sigma}=\sqrt{2}t_{\sigma}$ and $\bar{\sigma}=-\sigma$.
The new (higher order) GF on the right hand side of Eq.\
(\ref{gf1}) can also be determined from (\ref{gf2}), giving
\begin{widetext}
\begin{eqnarray}
(\omega-\varepsilon_d-U )\dlangle
c_{d\sigma}n_{d\bar{\sigma}};c^\dagger_{d\sigma} \drangle = \langle
n_{d\bar{\sigma}} \rangle
+\tilde t_{\sigma} \sum_k \left( \dlangle
c_{k\sigma}n_{d\bar{\sigma}};c^\dagger_{d\sigma} \drangle
- \dlangle c_{k\bar{\sigma}}c^\dagger_{d\bar{\sigma}}c_{d\sigma};c^\dagger_{d\sigma}
\drangle  +
\dlangle c^\dagger_{k\bar{\sigma}}c_{d\bar{\sigma}}c_{d\sigma};c^\dagger_{d\sigma}
\drangle \right) 
 \label{gf}.
\end{eqnarray}
\end{widetext}
\subsubsection{Coulomb blockade regime}
Although the EOM in Eq.\ (\ref{gf}) is exact, a
solution of the impurity GF requires a procedure to truncate and/or decouple the
higher order terms appearing on the right hand side of (\ref{gf}).  A solution that captures the
Coulomb blockade physics is given by the Hubbard-I approximation:\cite{14}
\begin{eqnarray} 
\dlangle c_{k\sigma}n_{d\bar{\sigma}};c^\dagger_{d\sigma} \drangle &
\simeq & \langle n_{d\bar{\sigma}}\rangle \dlangle
c_{k\sigma};c^\dagger_{d\sigma} \drangle \nonumber
\\
\dlangle
c_{k\bar{\sigma}}c^\dagger_{d\bar{\sigma}}c_{d\sigma};c^\dagger_{d\sigma}
\drangle & \simeq & \langle
c_{k\bar{\sigma}}c^\dagger_{d\bar{\sigma}}\rangle \dlangle
c_{k\sigma};c^\dagger_{d\sigma} \drangle \nonumber
\\
\dlangle
c^\dagger_{k\bar{\sigma}}c_{d\bar{\sigma}}c_{d\sigma};c^\dagger_{d\sigma}
\drangle & \simeq & \langle
c^\dagger_{k\bar{\sigma}}c_{d\bar{\sigma}}\rangle \dlangle
c_{k\sigma};c^\dagger_{d\sigma} \drangle , \label{HubbApprox}
\end{eqnarray}
which allows one to write
\begin{eqnarray}
G_{d\sigma}(\omega)\equiv \dlangle c_{d\sigma};c^\dagger_{d\sigma}
\drangle_{\omega}=\frac{G^{0}_{d\sigma}(\omega)}{1-G^{0}_{d\sigma}(\omega)\tilde{t}^2_{
\sigma}\tilde{g}(\omega)} \label{eqn11},
\end{eqnarray} 
where $G^{0}_{d\sigma}(\omega)=\frac{1-\langle n_{d\bar{\sigma}}
\rangle}{\omega-\varepsilon_d} +\frac{\langle n_{d\bar{\sigma}}
\rangle}{\omega-\varepsilon_d-U}$ is the local GF in the ``atomic"
approximation (the exact result for $t_{\sigma}=0$), and
$\tilde{g}(\omega)=\sum_{k}(\omega-\epsilon_k)^{-1}$ is the
non-interacting GF of the leads. The DOS of the system (proportional to the imaginary part of $G_{d\sigma}$)
contains two Hubbard peaks of width proportional to
$\Gamma_{\sigma}=\pi t^2_\sigma/D$, resulting in the broadening of the poles of
$G^{0}_{d\sigma}$. The spectral weights of these peaks are
controlled by the dot level occupancy with opposite spin, and caused by
the Coulomb interaction in the dot.  Notice that the SO-induced polarization of the QPC
results in different peak widths for the different spins.
The Hubbard-I approximation (\ref{HubbApprox}) is known to be valid
for a large $U/\Gamma$ ratio, when the Hubbard
subbands are well separated in energy scale. \cite{Mahan} It is the simplest
scheme which describes correlated electrons, although, since it ignores the Kondo effect, it is a
reasonable description only at temperatures higher than the Kondo scale ($T\gg T_K$--see next section).

\begin{figure}[h]
\includegraphics[width=3.5in]{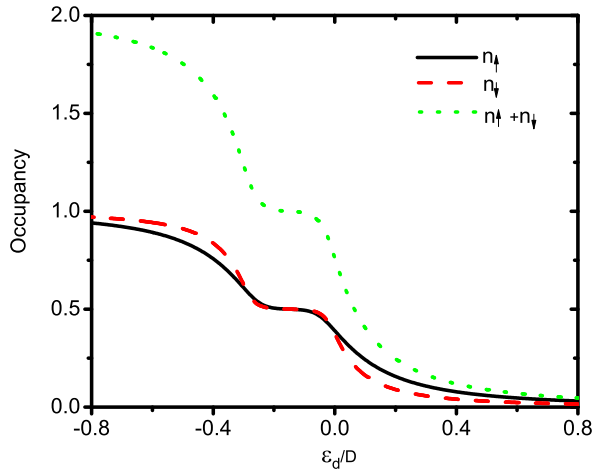}
\caption{(color online) Occupancies for spin-up $\langle
n_{d\uparrow}\rangle$, spin-down $\langle n_{d\downarrow}\rangle$
and total spin $\langle n_{d\uparrow}\rangle+\langle
n_{d\downarrow}\rangle$ vs.\ $\varepsilon_d$ at zero temperature.
Parameters used are $\Gamma_{\uparrow}=0.06$, $\Gamma_{\downarrow}=0.03$,
$U=0.3$, with $D=1$.  Notice asymmetry in $\langle n_{d\sigma} \rangle$ on each side of
the plateau. } \label{g2}
\end{figure}
\begin{figure}[h]
\includegraphics[width=1\linewidth]{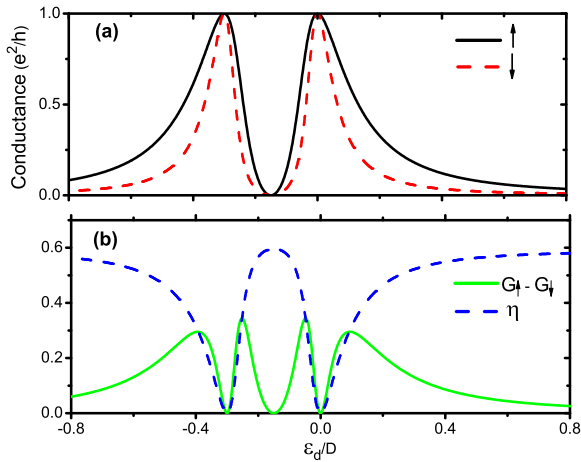}
\caption{(color online) Spin-dependent conductance and polarization as function of
 $\varepsilon_d$ at zero temperature. The other parameters are as in Fig.\ \ref{g2}. } \label{g3}
\end{figure}

The occupancies of spin $\uparrow$ and $\downarrow$ are
calculated self-consistently from the equation
\begin{eqnarray}
\langle n_{d\sigma}\rangle=\int
f(\omega) \left(-\frac{1}{\pi} \text{Im} \left[G_{d\sigma}(\omega ) \right] \right) d\omega \, , \label{occupancy}
 \label{eqn}
\end{eqnarray}
where $f(\omega)$ is the Fermi function. It is clear that when the QPCs are not polarizing,
$t_{\uparrow}=t_{\downarrow}$, the occupancy curves for
spin-up and down coincide and a plateau
of width $\sim U$ appears when the QD level moves below the Fermi level ($\varepsilon_d + U \gtrsim E_F \gtrsim \varepsilon_d$).
This situation changes when the QPCs are
polarized (see Fig.\ \ref{g2}), as the different $t_\sigma$ result in $\Gamma_{\uparrow} \neq
\Gamma_{\downarrow}$, which in turn produce different
$\langle n_{d\uparrow} \rangle$ and $\langle n_{d\downarrow} \rangle$, especially on both sides of the plateau.

The zero-bias conductance is calculated using a
Landauer formula generalized for interacting systems \cite{15}
\begin{eqnarray}
G_{\sigma}=\frac{e^2}{h}\Gamma_{\sigma} \int
 d\epsilon
 \frac{\partial f(\epsilon)}{\partial \epsilon} \, \text{Im}[G_{d\sigma}
(\epsilon) ] ,
\end{eqnarray}
for symmetric coupling of the leads for each spin. One can
also calculate the polarization factor
\begin{eqnarray}
\eta =\frac{G_{\uparrow}-G_{\downarrow}}{G_{\uparrow}+G_{\downarrow}} ,
\end{eqnarray}
which gives a measure of current polarization in the system.
Figure \ref{g3} shows the spin-dependent conductances
and polarization for the system in the Hubbard-I approximation.
As $\langle n_{d\uparrow}\rangle \neq \langle n_{d\downarrow}\rangle$ in general
(except at the particle-hole symmetry point, $\varepsilon_d = -U/2$),
the conductance per spin $G_{\sigma}$ are also different.
Notice that the spin-dependent conductance peaks are very asymmetric and
non-Lorentzian,
due to the peculiar behavior of the occupancies and their different up and down-spin
couplings.  As we consider here the case $\Gamma_{\uparrow}>\Gamma_{\downarrow}$,
one clearly sees that generally $G_{\uparrow} > G_{\downarrow}$ over the entire range of
$\varepsilon_d$ values. As a consequence, there is a net
up-spin polarization ($\simeq 60$\%) and conductance around the resonant peaks,
with the latter reaching $\simeq 0.3(e^2/h)$.

\subsubsection{Kondo regime}

To study the low-temperature behavior of the system within the EOM we need to consider
higher order GFs in Eq.\ (\ref{gf}). Using Lacroix's approach,
\cite{3} one can obtain relations for the three GFs on the right side of (\ref{gf}) as:
\begin{widetext}
\begin{eqnarray}
(\omega-\varepsilon_{k\sigma}) \dlangle
c_{k\sigma}n_{d\bar{\sigma}};c^{\dagger}_{d\sigma} \drangle &=& \langle
[c_{k\sigma}n_{d\bar{\sigma}};c^{\dagger}_{d\sigma}]_{+} \rangle +
\dlangle [c_{k\sigma}n_{d\bar{\sigma}};H];c^{\dagger}_{d\sigma}
\drangle \nonumber\\
 &=&\tilde t_{\sigma}
\dlangle n_{d\bar{\sigma}}c_{d\sigma};c^{\dagger}_{d\sigma}
\drangle+\tilde t_{\bar{\sigma}} \sum_{k'}[ \dlangle
c_{k\sigma}c^{\dagger}_{d\bar{\sigma}}c_{k'\bar{\sigma}};c^{\dagger}_{d\sigma}
\drangle- \dlangle
c^{\dagger}_{k'\bar{\sigma}}c_{d\bar{\sigma}}c_{k\sigma};c^{\dagger}_{d\sigma}
\drangle ],
\end{eqnarray}
\begin{eqnarray}
(\omega-\varepsilon_{d\sigma}+\varepsilon_{d\bar{\sigma}}-\varepsilon_{k\bar{\sigma}})
\dlangle c_{k\bar{\sigma}} c^{\dagger}_{d\bar{\sigma}} c_{d\sigma};c^{\dagger}_{d\sigma} \drangle&&= \langle
[c^{\dagger}_{d\bar{\sigma}}
c_{k\bar{\sigma}}c_{d\sigma};c^{\dagger}_{d\sigma}]_{+}\rangle +\dlangle
[c^{\dagger}_{d\bar{\sigma}}
c_{k\bar{\sigma}}c_{d\sigma};H];c^{\dagger}_{d\sigma} \drangle
\nonumber\\&& = \langle c^{\dagger}_{d\bar{\sigma}}c_{k\bar{\sigma}}
\rangle +\tilde t_{\bar{\sigma}} \dlangle
n_{d\bar{\sigma}}c_{d\sigma};c^{\dagger}_{d\sigma} \drangle+  \sum_{k'}[
 - \tilde t_{\bar{\sigma}} \dlangle
c^{\dagger}_{k'\bar{\sigma}}c_{k\bar{\sigma}}c_{d\sigma};c^{\dagger}_{d\sigma}
\drangle \nonumber\\&& + \tilde t_{\sigma}\dlangle
c^{\dagger}_{d\bar{\sigma}}c_{k\bar{\sigma}}c_{k'\sigma};c^{\dagger}_{d\sigma}
\drangle ],
\end{eqnarray}
and
\begin{eqnarray}
(\omega-\varepsilon_{d\sigma}-\varepsilon_{d\bar{\sigma}}+\varepsilon_{k\bar{\sigma}}-U)
\dlangle c^{\dagger}_{k\bar{\sigma}}
c_{d\bar{\sigma}}c_{d\sigma};c^{\dagger}_{d\sigma} \drangle&&= \langle
[c^{\dagger}_{k\bar{\sigma}}
c_{d\bar{\sigma}}c_{d\sigma};c^{\dagger}_{d\sigma}]_{+}\rangle + \dlangle
[c^{\dagger}_{k\bar{\sigma}}
c_{d\bar{\sigma}}c_{d\sigma};H];c^{\dagger}_{d\sigma} \drangle
\nonumber\\&& =
 \langle
c^{\dagger}_{k\bar{\sigma}}c_{d\bar{\sigma}}
\rangle-\tilde t_{\bar{\sigma}}\dlangle
n_{d\bar{\sigma}}c_{d\sigma};c^{\dagger}_{d\sigma} \drangle+\sum_{k'} [
\tilde t_{\sigma} \dlangle
c^{\dagger}_{k'\bar{\sigma}}c_{d\bar{\sigma}}c_{k'\sigma};c^{\dagger}_{d\sigma}
\drangle \nonumber\\&&  -\tilde t_{\bar{\sigma}} \dlangle
c^{\dagger}_{k\bar{\sigma}}c_{d\sigma}c_{k'\bar{\sigma}};c^{\dagger}_{d\sigma}
\drangle ] .
\end{eqnarray}
\end{widetext}

\noindent
Following the decoupling procedure in Ref.\ \onlinecite{3}, each
GF of the type $\dlangle A^{*}BC,D^{*} \drangle $ is
replaced by
\begin{eqnarray}
\approx \langle A^{*}B  \rangle
\dlangle C,D^{*} \drangle 
-\langle A^{*}C \rangle \dlangle B,D^{*} \drangle ,
\end{eqnarray}
resulting in an equation for the dot GF given by
\begin{widetext}
\begin{eqnarray}
G_{d\sigma}(\omega)&=&\left[U(\omega) - \langle
n_{d\bar{\sigma}}\rangle
-B_{\bar{\sigma}}(\omega)-B_{\bar{\sigma}}(\omega_1)\right] \nonumber\\
&&\times
\left\{U(\omega)[\omega-\varepsilon_{d\sigma}-\Sigma_{\sigma}(\omega)]+[B_{
\bar { \sigma }}(\omega)
+B_{\bar{\sigma}}(\omega_1)\right]\Sigma_{\sigma}(\omega)-
A_{\bar{\sigma}}(\omega)+ A_{\bar{\sigma}}(\omega_1)\}^{-1} \, ,
\label{gg1}
\end{eqnarray}
where $\Sigma_{\sigma}(x)=\sum_{k}| \tilde
t_{\sigma}|^2/(x-\varepsilon_{k\sigma})$,
$U(\omega)=[U-\omega+\varepsilon_{d\sigma}-\Sigma_{\sigma}(\omega)
+\Sigma_{\bar{\sigma}}(\omega)-\Sigma_{\bar{\sigma}}(\omega_1)]/U
$ and $\omega_1=-\omega +\varepsilon_{d\sigma}+U$. The functions
$A_{\sigma}(\omega)$ and $B_{\sigma}(\omega)$ are given by
\begin{eqnarray}
B_{\sigma}(\omega)&=&\frac{i}{2\pi} \int d\omega'
f(\omega') \left[
G_{d\sigma}(\omega')\frac{\Sigma_{\sigma}(\omega')-\Sigma_{\sigma}(\omega)}{\omega-\omega'-i\delta}
- G^{*}_{d\sigma}(\omega') \frac{\Sigma^{*}_{\sigma}(\omega')-\Sigma_{\sigma}(\omega)}{\omega-\omega'+i\delta} \right]
\nonumber \\
A_{\sigma}(\omega)&=&\frac{i}{2\pi} \int d\omega' f(\omega')
\left[ \left(1+G_{d\sigma}(\omega')\Sigma_{\sigma}(\omega') \right) \frac{\Sigma_{\sigma}(\omega')
-\Sigma_{\sigma}(\omega)}{\omega-\omega'-i\delta} -
\left(1+G^{*}_{d\sigma}(\omega')\Sigma^{*}_{\sigma}(\omega') \right)
\frac{\Sigma^{*}_{\sigma}(\omega')-\Sigma_{\sigma}(\omega)}{\omega-\omega'+i\delta}
\right] ,
\end{eqnarray}
\end{widetext} 
and have to be calculated self-consistently. In the limit $U
\rightarrow \infty$, the GF (\ref{gg1}) acquires a simpler form,
\begin{eqnarray}
G_{d\sigma}(\omega)=\frac{ 1 - \langle n_{d\bar{\sigma}}\rangle
-B_{\bar{\sigma}}(\omega)
}{\omega-\varepsilon_{d\sigma}-(1-B_{\bar{\sigma}}(\omega))
\Sigma_{\sigma}(\omega)- A_{\bar{\sigma}}(\omega)} . \label{eqn1}
\end{eqnarray}

\begin{figure}[h]
\includegraphics[width=1\linewidth]{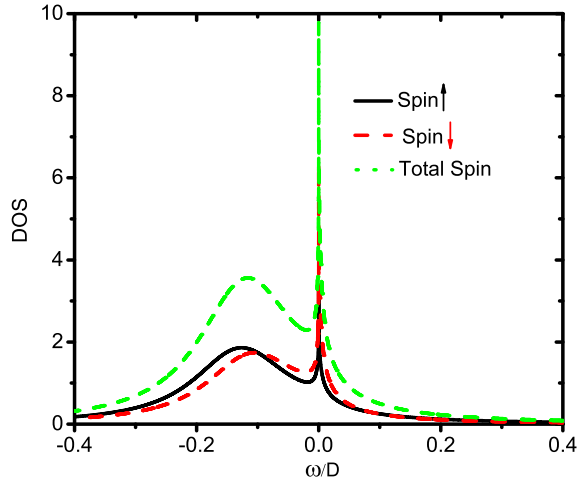}
\caption{(color online) Density of states in QD as a function of $\omega$ in the Kondo regime. Parameters
used are $T=10^{-7}$,
$\varepsilon_d=-0.136 $, $\Gamma_{\uparrow}=0.06$ and
$\Gamma_{\downarrow}=0.03$ ($D=1$).  Sharp feature near the Fermi level ($\omega=0$) is the signature
of the Kondo screening, although asymmetric here for the different spin species. } \label{anh}
\end{figure}

\begin{figure}[h]
\centerline{\resizebox{3.5in}{!}{
\includegraphics{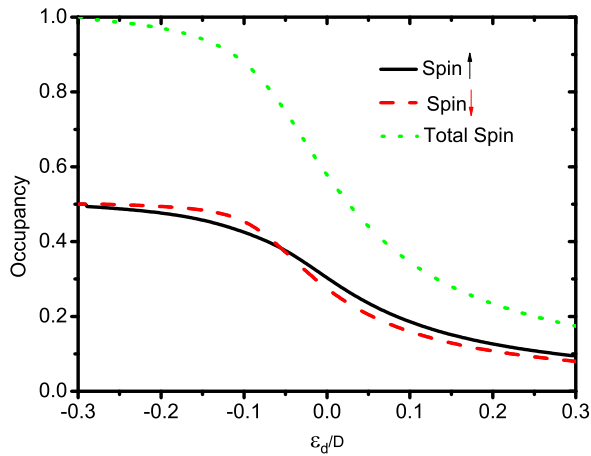}}}
\caption{(color online) Occupation in the QD as function of $\epsilon_d$. Parameters used
as in Fig.\ \ref{anh}. }
 \label{charge}
\end{figure}

In the following, we solve numerically for the spectral function (DOS) in the
Kondo regime by the self-consistent iteration of Eqs.\ (\ref{eqn}) and (\ref{eqn1}).
Figure \ref{anh} shows the DOS vs.\ $\omega$
at low temperature ($T= 10^{-7}$) for both spin
orientations (and total), when the QD electron level $\varepsilon_d$ is taken to be
at $-0.136$.  The three curves exhibit Kondo resonance peaks near the Fermi
level ($\omega \simeq 0$), in addition to a much broader peak at
$\omega \simeq \varepsilon_d$. \cite{3}  Since
the hybridization between leads and QD are spin dependent in this case, the
DOS clearly splits into two different components for
spin up and down. The DOS for spin down is shifted upwards in energy with
respect to the spin up component, resulting in a lower occupancy for the down
spin.   Figure
\ref{charge} shows indeed the occupancy curves for $\langle n_{d\uparrow} \rangle$,
$\langle n_{d\downarrow} \rangle$ and total $\langle n \rangle$ vs.\ $\varepsilon_d$ at a given
temperature ($T= 10^{-7}$).
Qualitatively similar to the results in the Coulomb blockade regime, the
occupancy curves show that $\langle n_{d\uparrow} \rangle > \langle n_{d\downarrow} \rangle$
for $\omega \gtrsim -0.05$, while the relation is reversed for smaller $\omega$ values,
reflecting the asymmetry introduced by the polarizing QPCs, through $\Gamma_\uparrow$
and $\Gamma_\downarrow$.  Notice also that the plateaus at 0.5
are not as well defined here, due to the enhanced spin and
charge fluctuations in the Kondo regime.
The spin-orbit effect can be understood qualitatively as arising from a shift
in the dot level (as well as depending on the occupancy factors): since the effective
level position of the electron
with spin $\sigma$ is given by $\varepsilon_{d\sigma}\approx
\varepsilon_{d}+ \text{Re} \Sigma^{'}_{\sigma}(\omega)$,  
where $\Sigma^{'}_{\sigma}(\omega) \propto t_{\sigma}^2$ is the self-energy, and thus
naturally causes the spin-dependent occupancy seen in the figure.

\begin{figure}[h]
\includegraphics[width=1\linewidth]{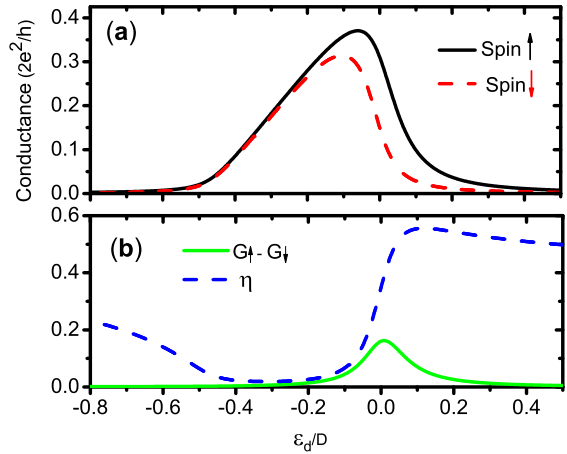}
\caption{(color online) (a) Conductance, and (b) polarization and
difference between spin up and down conductance, as functions of
$\epsilon_d$. Parameters as in
 Fig.\ \ref{anh}. }
 \label{conduct}
\end{figure}

Figure \ref{conduct}(a) shows the spin-dependent conductance curves
 $G_{\sigma}$ vs.\ $\varepsilon_d$. Several features are noteworthy. As  $\varepsilon_d$
changes, the spin-dependent conductances exhibit the
anticipated peaks at low temperature, with spin up conductance dominating
(naturally, as $\Gamma_\uparrow > \Gamma_\downarrow$). Figure \ref{conduct}(b) shows the
difference between spin up and down conductances as well as the spin
polarization vs.\ $\varepsilon_d$.  In this regime, the
difference between the conductance for the
two spin orientations reaches $\simeq 0.4 (e^2/h)$.
Correspondingly, the net spin polarization reaches $\eta \simeq 60$\%.

\begin{figure}[tb]
\includegraphics[width=1\linewidth]{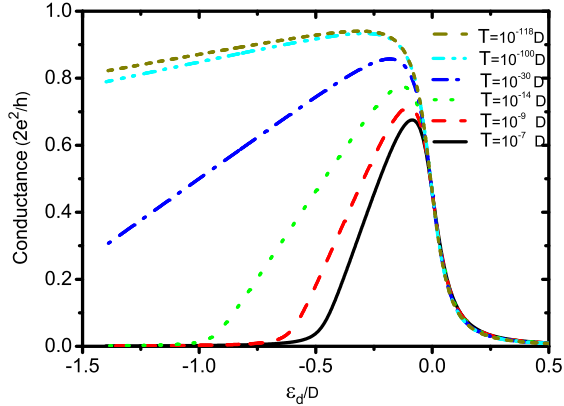}
\caption{ (color online) Total conductance as function of $\varepsilon_d$ for several
temperatures $T$.  Notice that $T_K$ is strongly suppressed for more negative $\varepsilon_d$ values,
which lowers the conductance at a given $T$. Parameters as in  Fig.\ \ref{anh}. }
\label{conductT}
\end{figure}

Let us now analyze in more detail the effect of temperature on the conductance
of the system and especially its drop for $\varepsilon_d \ll 0$. The conductance curves in Fig.\ \ref{conduct}(a) are highly
asymmetric about the Fermi energy and vanish rapidly away from it.
This vanishing for very negative values of $\varepsilon_d$ is
due to the the Kondo temperature, $T_K$, becoming smaller than
the temperature of the system.
Figure \ref{conductT} presents the total zero bias
conductance as function of $\varepsilon_d$  for several values of
temperature. As $T$ is lowered, the total conductance
increases for a given $\varepsilon_d$, and the width of the
conductance peak increases in $\varepsilon_d$. This explicitly
reflects the existence of the Kondo resonant peak in the spectral
function, and how the system will reach the unitary limit of
conductance for $T=0$ for $\varepsilon_d$ well below the Fermi level.

Our discussions above are applied to the case of infinite $U$, where the EOM method
gives qualitatively accurate results in the Kondo regime.  Extension of this approach
for finite $U$ is known to be problematic, including its failure to exhibit a Kondo
resonance at the particle-hole symmetry point
($\varepsilon_d=-U/2$). \cite{18}
To carry out our
study in the finite $U$ case, we use instead the essentially exact
numerical renormalization group approach, as we discuss in the following section.

\subsection{NRG results for finite U} \label{NRGsect}

\begin{figure}[h]
\includegraphics[width=1\linewidth]{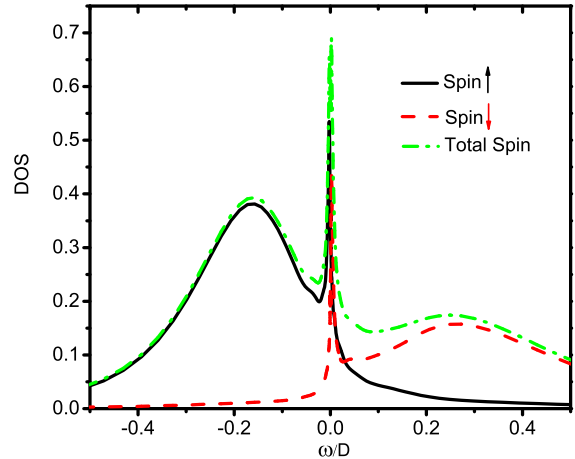}
\caption{(color online) Density of states in the QD as function of energy
obtained from NRG calculations.  Polarizing QPCs generate effective splitting
of the DOS for different spins. System parameters used are
$U=0.5$, $\Gamma_{\uparrow}=0.06$, $\Gamma_{\downarrow}=0.03$, $D=1$, and $T=0$.}
\label{NRG1}  
 \end{figure}

For the finite-$U$ case we study the spin polarized conductance
using the standard numerical renormalization group
approach.\cite{Wilson75,costi} In this case, unlike
the previous infinite-$U$ case, the processes involving double
occupied states are naturally present in the dynamics of the system,
and allow for a reliable description of the low-energy behavior.
We set $U=0.5$ and $T=0$. Figure \ref{NRG1} depicts the local
density of states calculated with NRG for the same system
parameters as before,
$\Gamma_{\uparrow}=0.06$, $\Gamma_{\downarrow}=0.03$, and
with $\epsilon_d=-0.2$.
This value of $\epsilon_d$ corresponds to a situation where the
system is away from the particle-hole symmetric point ($=-U/2$),
more suitable to compare to the previous infinite-$U$
calculation (where the system never reaches the {\it p-h} symmetry point).
Notice that there is a strong spin asymmetry in the DOS; on the
negative side of the $\omega$-axis, the DOS for spin up
(solid curve) presents a peak near $-0.2$, corresponding to the
energy of the local bare orbital $\varepsilon_d$, slightly shifted by the real part
of the proper self-energy. On the positive side of the $\omega$-axis, the peak near
$\epsilon_d+U$ would result in the succeeding CB peak, which
appears only for the spin-down DOS (dashed curve). In this regime as well, the fact that
$\Gamma_\uparrow > \Gamma_\downarrow$ favors the spin-up occupancy to the
detriment of the spin-down occupancy.
One also notices  the important
peaks close to the Fermi level, signature of the Kondo effect, which are
slightly split away and
suppressed by a seemingly effective magnetic field induced by the
spin-asymmetric coupling to the leads. This phenomenon is akin to the
suppression discussed by Martinek {\it et al.,}\cite{Martinek} in the context of a QD coupled to
ferromagnetic leads.  Notice, however, that no external magnetization is present
in our system and that the polarization is only arising from the QPCs and due
to the lateral SO interaction.

\begin{figure}[h]
\includegraphics[width=1\linewidth]{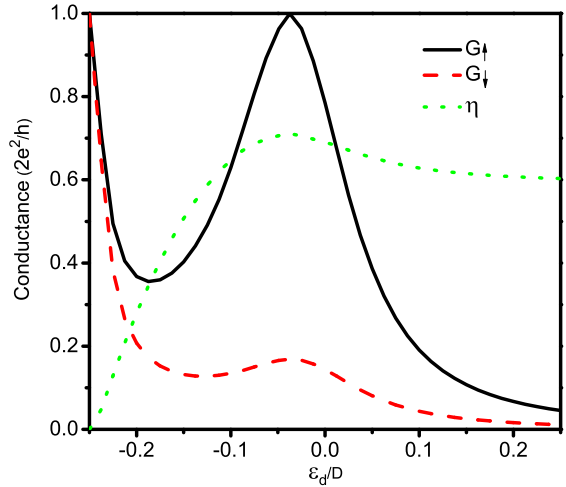}
\caption{(color online) Total conductance as function of $\varepsilon_d$
obtained from NRG calculations. Same system parameters as in Fig.\ \ref{NRG1}. }
 \label{NRG2}
\end{figure}

Analogously to the infinite-$U$ case, the spin-asymmetry discussed
above induces spin polarized transport in the system. In
Fig.\ \ref{NRG2} we show the conductance as a function of
$\varepsilon_d$ for the same parameters as Fig.\ \ref{NRG1}. Notice
that away from the {\it p-h} symmetric point ($\varepsilon_d=-0.25$)
the conductance for spin up is much larger than for spin down,
resulting in a sizeable polarization ($\eta \simeq 70$\%), as shown by
the (green) dotted curve. At the {\it p-h} symmetric
point the conductance for both spins reaches the unitary limit and
$\eta\rightarrow 0$; this is consistent with the restoration of the Kondo state
of the system at the {\it p-h} point for ferromagnetic leads. \cite{Sindel}

\section{Summary} \label{summ}

In summary, we have investigated the spin-dependent transport
properties of quantum dot structures with polarizing quantum point contacts.
We have shown that as QPCs can generate
finite spin-polarized currents, due to the combination of
lateral and perpendicular spin-orbit interactions, they also induce
current polarization in quantum dots made with these QPCs. Using
equation-of-motion techniques and numerical renormalization group
calculations, we obtained the electronic Green's function, conductance
and spin polarization in different parameter regimes. Our results
demonstrate that both in the Coulomb blockade and Kondo regimes, the quantum
dot exhibits non-zero
spin-polarized conductance, even when the injection is unpolarized and
there are no applied magnetic fields. The spin-dependent coupling is shown to
give rise to nontrivial effects in the density of states of the
single QD, resulting in strong modification of the charge distribution
in the system. Most importantly, these effects are controllable by lateral
gate voltages applied to the QPCs, and together with the ability to create
quantum dots, they
provide a new approach for exploring spintronic devices, spin
polarized sources and spin filters.

\section{Acknowledgements}
We thank helpful discussions with P. Debray and N. Sandler, as well as financial support from
CNPq, CAPES, and FAPEMIG in Brazil, and NSF-PIRE, and NSF-MWN/CIAM in the US.

\end{document}